# Hierarchical auxetic and isotropic porous medium with extremely negative Poisson's ratio

Maryam Morvaridi[1], Giorgio Carta[2], Federico Bosia[1], Antonio S. Gliozzi[1], Nicola M. Pugno[3,4], Diego Misseroni[3,(*)], Michele Brun[2,(**)]

[1]*Department of Applied Science and Technology, Politecnico di Torino, Corso Duca degli Abruzzi 24, 10129 Torino, Italy*

[2]*Dipartimento di Ingegneria Meccanica, Chimica e dei Materiali, Università di Cagliari, Piazza d'Armi, 09123 Cagliari, Italy*

[3] *Laboratory of Bio-inspired, Bionic, Nano, Meta Materials & Mechanics, Department of Civil, Environmental and Mechanical Engineering, University of Trento, Via Mesiano 77, 38123 Trento, Italy*

[4] *School of Engineering and Materials Science, Queen Mary University of London, Mile End Road, London E1 4NS, UK*

(*) diego.misseroni@unitn.it; (**) mbrun@unica.it

**Abstract**

We propose a novel two-dimensional hierarchical auxetic structure, consisting of a porous medium in which a homogeneous matrix includes a rank-two set of cuts characterised by different scales. The six-fold symmetry of the perforations makes the medium isotropic in the plane. Remarkably, the mesoscale interaction between the first- and second-level cuts enables the attainment of a very small value of the Poisson's ratio, close to the minimum reachable limit of -1. The effective properties of the hierarchical auxetic structure are determined numerically, considering both a unit cell with periodic boundary conditions and a finite structure containing a large number of repeating cells. Further, results of the numerical study are validated experimentally on a polymeric specimen with appropriately arranged rank-two cuts, tested under uniaxial tension. We envisage that the proposed hierarchical design can be useful in numerous engineering applications exploiting an extreme auxetic effect.

**Introduction**

Auxetic materials are characterised by the unconventional property of possessing an effective negative Poisson's ratio, so that they expand (contract) transversally when stretched (compressed) longitudinally.

Named "auxetic" after Evans [1], these media owe their special behaviour mainly to their microstructure rather than to their chemical composition. Hence, auxeticity has been observed at different scales, from macro- to nano-dimensions. Apart from examples of natural materials with an intrinsic negative Poisson's coefficient [2-5], auxetic media are generally artificially-made systems whose microstructure is designed by exploiting different geometries and mechanisms: re-entrant unit cells [6-9], star-shaped inclusions [10, 11], chiral configurations [12-15], double arrowhead honeycombs [16], perforations and cuttings [17-21], rotating rigid units [22], lattices [23-25] and elastic instabilities [26, 27]. Alternative approaches to design auxetic media are presented in the reviews [28-32]. Some auxetic systems are also characterised by a negative value of the coefficient of thermal expansion, implying that they shrink when subjected to an increase in temperature [33-35]. Additionally, it has been shown that it is also possible to achieve a smooth transition through a wide range of negative and positive Poisson's ratios by using an origami cell that morphs continuously between a Miura mode and an eggbox mode [36-38].

The increasing interest of the scientific community in auxetic metamaterials is due to their enhanced mechanical properties with respect to those of conventional materials, including higher indentation resistance [39] and impact energy absorption abilities [40, 41], improved fatigue performance [42] and pull-out strength [43], as well as the possibility to bend with synclastic curvature [44]. For these reasons, auxetic metamaterials have tremendous potential in many fields, particularly in the aviation industry, e.g. for aircraft design [43, 45, 46], in sports applications for enhanced comfort and protection [47], in electronics to increase electric power output [48] and in biomedical engineering for the design of novel types of stents [49-51] and orthopedic implants [52]. On the other hand, zero Poisson's ratio materials have been shown to be useful in other applications, for instance, in the design of morphing aircraft [53-55].

Hierarchical structures are widely exploited in natural materials [56], and in bioinspired artificial materials [57] to enhance mechanical properties. The simultaneous presence of multiple length scales, together with material heterogeneity, has been shown to allow the simultaneous optimization of strength and toughness [58, 59], but also to enable the improvement of other mechanical characteristics such as adhesion [60, 61], friction [62], or to achieve band gap engineering in dynamics [63, 64].

In this paper, we investigate the auxetic behaviour of a porous hierarchical medium, consisting of a homogeneous elastic material containing two classes of perforations, characterised by two different length scales and exhibiting a six-fold symmetry, which makes the medium isotropic [65, 66]. First, by numerically studying the periodic elementary cell, we demonstrate that, for some ratios between the lengths of the two classes of cuts, the effective Poisson's ratio approaches the limit of –1. Second, we verify the results of the periodic analysis by determining the response of a finite model, containing a large number of periodic cells, under uniaxial loading. Subsequently, we validate the numerical results by experimentally testing a specimen with hierarchical cuts in uniaxial tension. Finally, we provide some analysis and concluding remarks.

**Numerical model**

We introduce a 2-D hexagonal periodic unit cell with oriented cuts, as shown in Fig. 1a. The cuts are all rotated by the same relative angle $\Theta$ ($\Theta$-$\pi$/6 with respect to the intersected edge of the hexagonal unit cell of size $l$). Each cut is of length $a$ and width $b$, which is also the diameter of the rounded extremities (in consideration of a practical realization). The proposed design leads to a periodic pattern with either six-fold or three-fold symmetry, similar to previously considered designs providing an isotropic auxetic response [19, 20].

Next, we construct a second-rank geometry by adding smaller cuts to the first, arranged periodically in a hexagonal pattern of size $l'$ and in turn oriented at the same angle $\Theta$ (Fig. 1b). This entails the introduction of a second characteristic size scale in the system, leading to a so-called "hierarchical" geometry. The transposition of these 2-D geometries in real 3-D structures is implemented in thin sheets of thickness $t$, shown in the bottom panels of Fig.1.

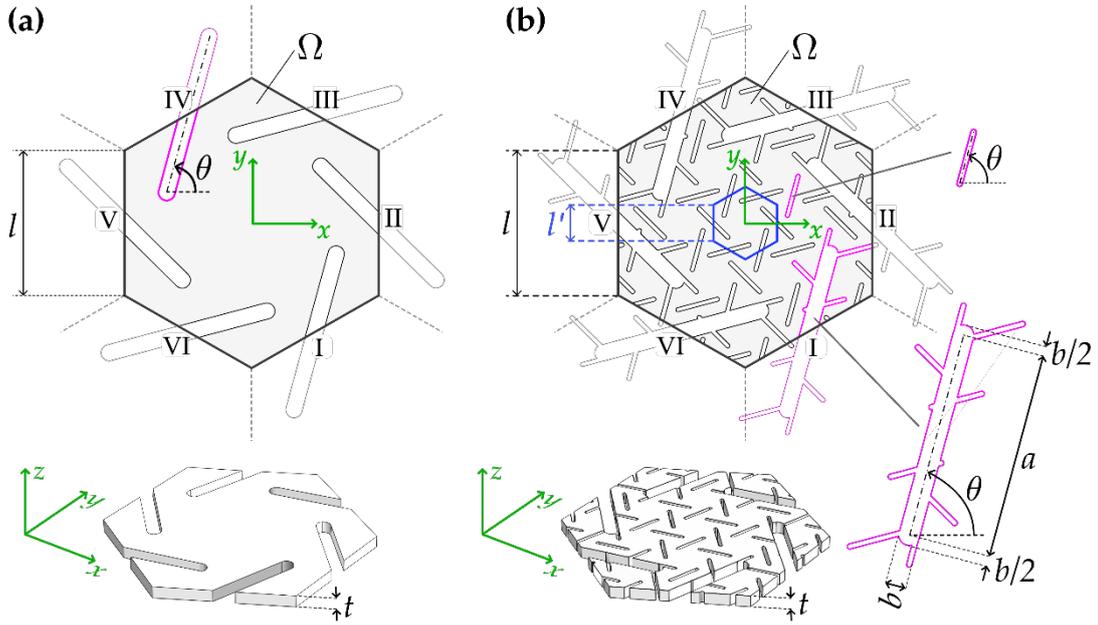

*Fig. 1: Top and isometric views of the considered unit cells: a) non-hierarchical rank-one hexagonal unit cell; b) hierarchical rank-two unit cell, with additional second-level cuts.*

We numerically evaluate the quasi-static behaviour under uniaxial tension of the two proposed 2-D geometries. For both of them we derive the "effective" properties as functions of the geometric parameters. This is accomplished by performing a Finite Element (FE) analysis using the commercial package *COMSOL Multiphysics*® (version 5.6). A state of plane stress is assumed in the numerical simulations, and periodic boundary conditions are applied at the edges of the unit cell. Geometrical parameters are taken as follows: cut length $a = 10$ mm, cut width $b = 1$ mm, lengths of the hexagonal cells $l = 9.00$ mm and $l' = 2.25$ mm, Young's modulus $E = 2.285$ GPa and Poisson's coefficient $\nu = 0.370$.

The effective elastic properties of the structure are derived by applying macroscopic longitudinal strains in the horizontal ($\varepsilon_{xx} = 10^{-4}$) and vertical ($\varepsilon_{yy} = 10^{-4}$) directions and the macroscopic shear strain $\varepsilon_{xy} = 10^{-4}$. Each of these macroscopic strains is applied to the unit cell using periodic boundary conditions. Specifically, periodic boundary conditions link displacements on edges having opposite outward normal unit vectors. Referring to Fig. 1, the periodic boundary conditions satisfy the relations $u_i^{(\alpha)} - u_i^{(\alpha+\text{III})} = \varepsilon_{ij}\left(X_j^{(\alpha)} - X_j^{(\alpha+\text{III})}\right)$, where $\alpha = $ I, II, III while $u_i$ are the displacements, $X_j$ the positions and $i, j = x, y$. In numerical computations, additional constraints are imposed to prevent rigid-body motions. The corresponding macroscopic Cauchy stress components $\sigma_{xx}$, $\sigma_{yy}$ and $\sigma_{xy}$ are evaluated numerically as average values of the corresponding local components $s_{xx}$, $s_{yy}$ and $s_{xy}$ on the unit cell domains of the models, namely:

$$\sigma_{ij} = \frac{1}{|\Omega|}\int_\Omega s_{ij}\, d\mathbf{x} = (1-p)\sigma_{ij}^{(S)} + p\sigma_{ij}^{(P)} = (1-p)\sigma_{ij}^{(S)} \quad (ij = xx, yy, xy), \tag{1}$$

where $\Omega$ is the unit cell domain, $\sigma_{ij}^{(S)}$ and $\sigma_{ij}^{(P)} = 0$ are the average stresses in the solid (S) and porous (P) phases, respectively, and $p$ is the porosity.

From these values, it is possible to estimate the effective Young's modulus $E_{eff}$ and Poisson's ratio $\nu_{eff}$ for plane stress conditions. From Hooke's law, the effective properties of an isotropic material are given by (see Appendix A)

$$E_{eff} = \frac{\sigma_{xx}^2 - \sigma_{yy}^2}{\varepsilon_{xx}\sigma_{xx} - \varepsilon_{yy}\sigma_{yy}}, \tag{2}$$

$$\nu_{eff} = \frac{\varepsilon_{xx}\sigma_{yy} - \varepsilon_{yy}\sigma_{xx}}{\varepsilon_{xx}\sigma_{xx} - \varepsilon_{yy}\sigma_{yy}}, \tag{3}$$

$$\mu_{eff} = \frac{\sigma_{xy}}{2\,\varepsilon_{xy}} = \frac{E_{eff}}{2(1+\nu_{eff})}. \tag{4}$$

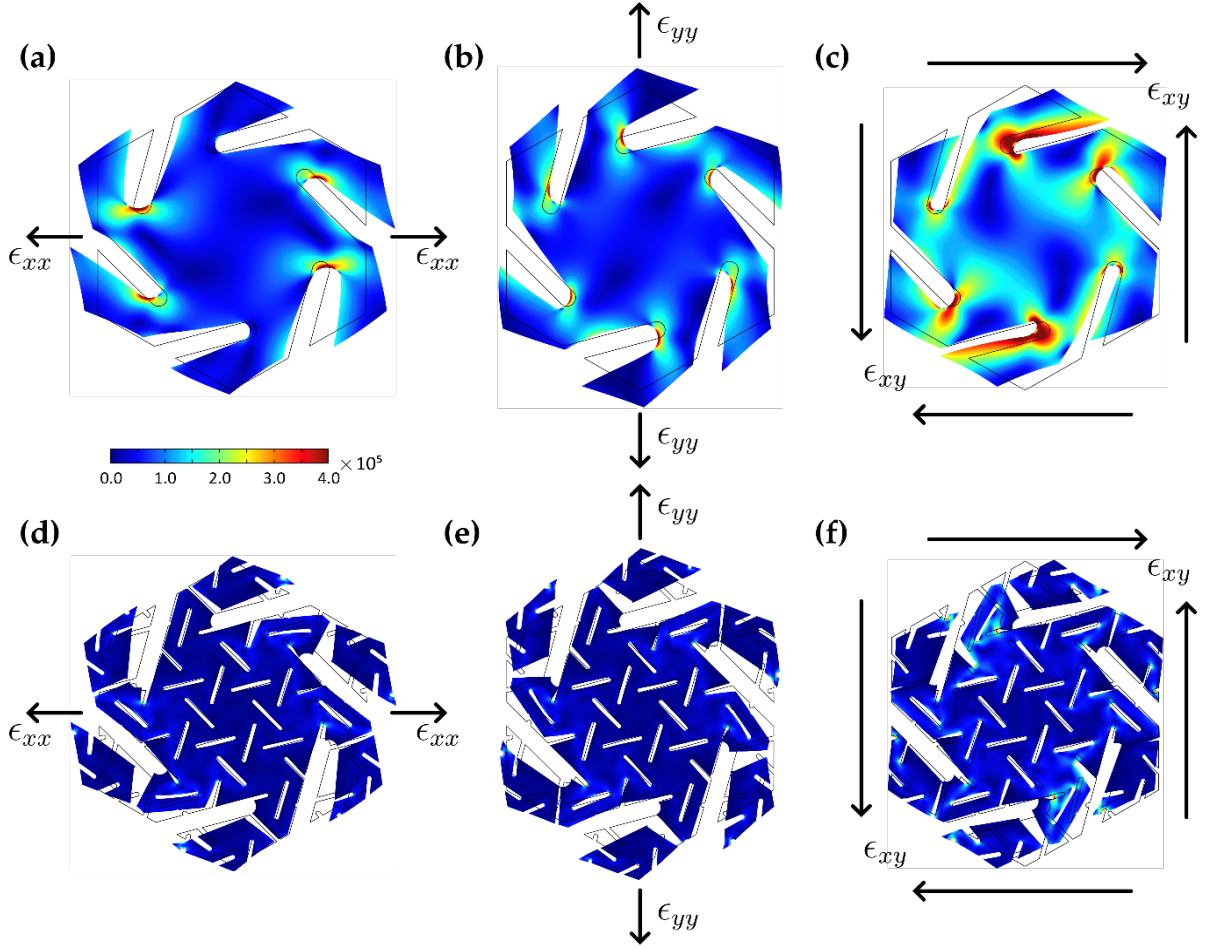

*Fig. 2: FE simulations on (a-c) non-hierarchical and (d-f) hierarchical unit cells. (a), (d): normal strain applied in the horizontal direction $\varepsilon_{xx} = 10^{-4}$ ; (b), (e): normal strain applied in the vertical direction $\varepsilon_{yy} = 10^{-4}$ ; (c), (f): shear strain $\varepsilon_{xy} = 10^{-4}$. Resulting von Mises stress distributions are represented in colour scale, where the values are given in N/m². For comparison purposes the fields have been plotted with the same colour scale.*

The equations above provide the homogenised properties of the 2-D linear elastic material, characterised by six-fold or three-fold symmetry. The FE simulations, performed with the three imposed strain components indicated above, confirm the isotropic behaviour of the structure, for both the hierarchical and non-hierarchical designs (the detailed analysis is reported in Appendix A). The local von Mises stress distributions $s_{VM}$ for both the non-hierarchical and hierarchical models, under the three strain conditions, are illustrated in Fig. 2. The local stress field components $s_{xx}$, $s_{yy}$ and $s_{xy}$ are also shown in Appendix B.

The effective Young's modulus and Poisson's ratio obtained for the non-hierarchical geometry are $E_{eff} = 201$ MPa and $\nu_{eff} = -0.55$ (NHEP: Non-Hierarchical Effective Properties), confirming the auxetic behaviour of this design. In the case of the hierarchical geometry, taking $l = 4l'$, we obtain $E_{eff} = 2.92$ MPa and $\nu_{eff} = -0.92$, i.e. there is a considerable reduction in the stiffness, a remarkable increase in the auxetic behaviour, with an effective Poisson's ratio tending to the minimum reachable value -1. Thus, as assumed, the hierarchical structure can yield a considerable variation in the mechanical properties of architected structures, with the possibility of achieving extreme values. Notice that, despite a reduction in effective Young's modulus between the hierarchical and non-hierarchical case, stress levels remain comparable, due to high stress concentrations at the tips of the cuts; the maximum local von Mises stresses are reported in Tables A.1 and A.2 in Appendix A.

These properties are strongly dependent on the choice of the microstructural geometrical parameters, particularly the so-called "hierarchical ratio" between the characteristic size of each rank of cuts. Therefore, we evaluate the variation of the mechanical properties of the structure ($E_{eff}$ and $\nu_{eff}$) with respect to the hierarchical ratio $s = l'/l$, for values ranging between 1/18 and 1/2. For each hierarchical geometry, the centre of the unit cell of the rank-one system of cuts coincides with the centre of the central unit cell of the rank-two system of cuts in order to enforce the overall six-fold symmetry.

Results are detailed in Table 1 and shown graphically in Fig. 3, together with the effective properties of the non-hierarchical structure (NHEP - horizontal blue lines) and of the "Matrix with effective properties" (MEP - horizontal red lines). In the computations, the hierarchical structure corresponding to MEP has been implemented as a non-hierarchical geometry, and the local properties of the matrix phase have been taken as the effective properties of the non-hierarchical system (NHEP). When the hierarchical ratio $s \to 0^+$, separation of length scales holds between the local perturbations of the fields (stress and strain) induced by the two ranks of cuts. In such a case, it is justified to model the matrix phase as a homogeneous material with effective properties instead of introducing rank-two cuts.

*Table 1: Numerically calculated effective Young's modulus, Poisson's ratio and porosity for varying hierarchical ratio $s = l'/l$. In (*) a supercell made of 2x2 cells was analysed to include an integer number of rank-two cells of size $l'$.*

| $s$ | $E_{eff}$ (MPa) | $\nu_{eff}$ | Porosity (%) |
|---|---|---|---|
| 1/2 | 6.33 | -0.03 | 27 |
| 1/3 | 36.5 | -0.22 | 24 |
| 1/4 | 2.92 | -0.92 | 28 |
| 1/5 | 11.3 | -0.60 | 25 |
| 1/5.5 (*) | 16.9 | -0.50 | 29 |
| 1/6 | 4.22 | -0.85 | 23 |
| 1/6.5 (*) | 16.1 | -0.58 | 28 |
| 1/7 | 12.1 | -0.62 | 24 |
| 1/8 | 19.4 | -0.61 | 29 |
| 1/9 | 18.8 | -0.55 | 26 |
| 1/12 | 17.2 | -0.58 | 28 |
| 1/15 | 14.2 | -0.63 | 28 |
| 1/18 | 15.7 | -0.62 | 27 |
| $0^+$ | 17.6 | -0.63 | 29 |

Fig. 3 confirms that by decreasing the hierarchical ratio $s$, both the Young's modulus and the Poisson's ratio converge to the limiting values corresponding to MEP ($E_{eff} = 17.64$ MPa and $\nu_{eff} = -0.632$). On the other hand, for larger ratios $s \in [1/6, 1/2]$ the numerical results show large oscillations of the elastic properties. Remarkably, the Poisson's ratio is minimised at the mesoscale, where the two scales associated with the two ranks of cuts strongly interact locally. Combining constructively the effect of hierarchy and scale interaction, the minimum values of Poisson's ratio are found for $s = 1/6$ and $s = 1/4$, reaching in the latter case the value $\nu_{eff} = -0.92$, that is very close to the thermodynamic limit for constitutive stability $\nu_{eff} = -1$.

Clearly, the introduction of the second rank of cuts increases the effective porosity and, as expected, causes a decrease in Young's modulus with respect to the non-hierarchical model. Notice that the exceptional value of $\nu_{eff}$ does not depend on the particularly small value of $E_{eff} = 2.92$ MPa for this geometry: calculations for a non-hierarchical unit cell with $E = E_{eff}$ yield $\nu_{eff} = -0.55$.

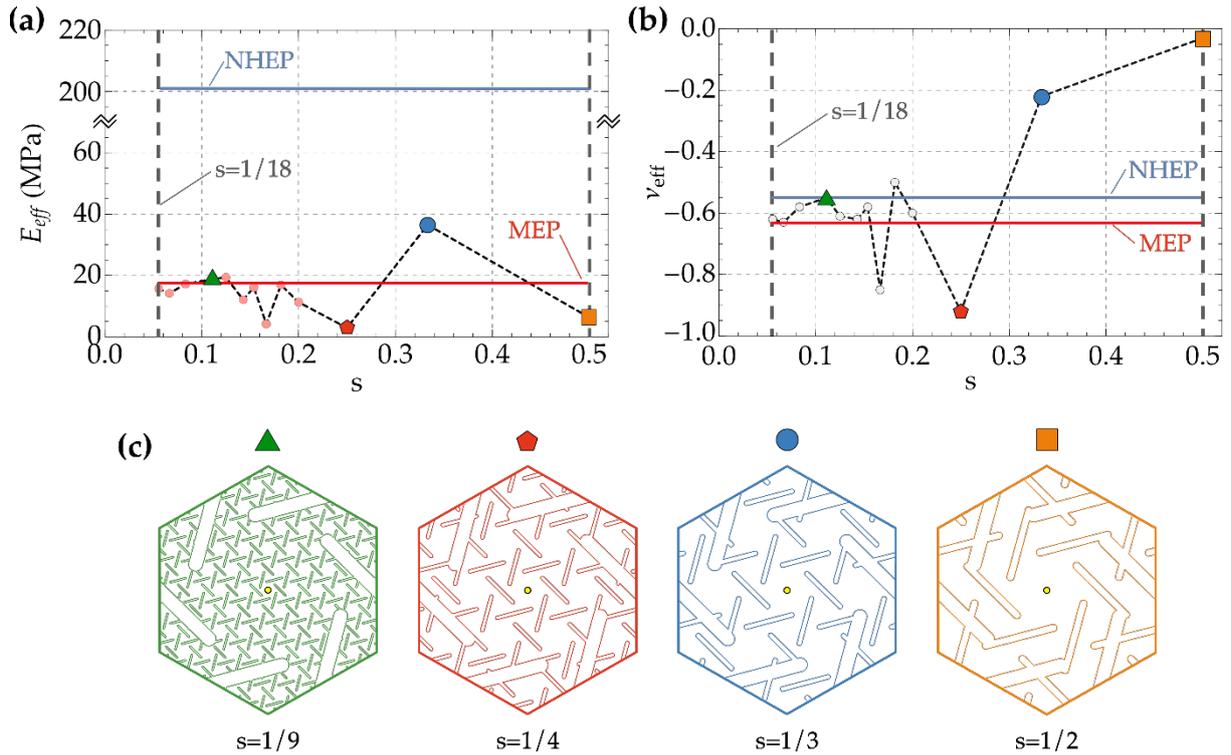

*Fig. 3: (a) Effective Young's modulus (note the jump in the scale on the vertical axis) and (b) effective Poisson's ratio of hierarchical-auxetic material as function of hierarchical ratio s. Four of the considered hierarchical auxetic structures are also shown in panel (c).*

Concerning the potential benefits of a multiple-rank microstructure, we note that the results for the hierarchical ratio *s* = 1/4 are already close to the limiting value -1. Additional computations under the separation of length scale hypothesis, i.e. iterating the MEP scheme, and reported in Table 2. Their show that additional hierarchical levels lead to a small decrease in Poisson's ratio at the cost of a strong reduction of Young's modulus, which is accompanied by additional technological difficulties in specimen manufacture.

*Table 2: Effective Young's modulus and Poisson's ratio for multiple-rank microstructures and $s \to 0^+$.*

| Rank | $E_{eff}$ (MPa) | $\nu_{eff}$ |
|------|-----------------|-------------|
| 1    | 2285            | 0.370       |
| 2    | 201             | -0.551      |
| 3    | 17.6            | -0.632      |
| 4    | 1.55            | -0.639      |
| 5    | 0.136           | -0.640      |
| 6    | 0.012           | -0.640      |

To further verify the auxetic behaviour of the proposed design, we study a finite two-dimensional model of dimensions 280×600 mm, including numerous unit cells, under plane stress conditions. With this model, we check that the effective values of Young's modulus and Poisson's ratio at this scale coincide with those estimated for a single periodic unit cell. We consider the structure corresponding to a hierarchical ratio $s$ = 1/4. The horizontal and vertical displacement fields have been evaluated under uniaxial tension applied in the vertical direction, and leaving free boundary conditions on lateral sides of the specimen (Fig. 4). The effective Young's modulus and Poisson's ratio are evaluated using Eqs. (2) and (3), where the effective strain and stress components are computed averaging the corresponding local fields in the domain enclosed in the rectangle R indicated by a black line in Fig. 4a. Additionally, the effective Poisson's ratio is also determined as the (negative) ratio between the average strains in the horizontal ($\varepsilon_{xx}$) and vertical ($\varepsilon_{yy}$) directions. These last, have been evaluated from the average horizontal displacements computed on the vertical lines connecting points 1 – 6 and 3 – 8 and from the average vertical displacements calculated on the horizontal lines connecting points 1 – 3 and 6 – 8, respectively (specified in Fig. 4a); this second approach is implemented to mimic the experimental procedure detailed below (see Fig. 5c). The two approaches yield values of Poisson's ratio which are the same up to the second decimal digit. The resulting effective Young's modulus and Poisson's ratio values are $E_{eff} = 2.9$ MPa and $v_{eff} = -0.90$, showing a very good agreement with previously calculated values for single periodic unit cells.

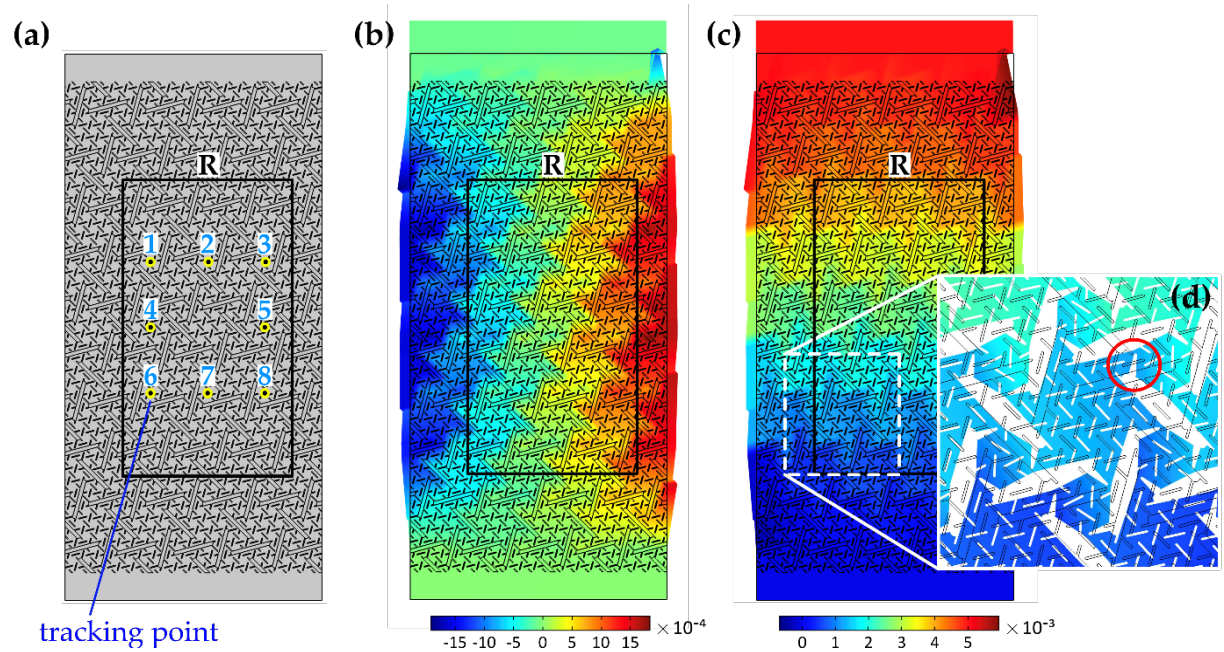

*Fig. 4: (a) Considered large-scale geometry. The rectangle R indicates the region where average values of stress and strain are computed for the evaluation of the effective properties. (b) Horizontal and (c) vertical displacements (given in meters) for an applied vertical displacement at the top side of the sample. The displacement components are given in colour scales on the deformed sample. d) Close-up of unit cell deformation.*

The contour plots in Figs. 4b and 4c clearly show the material expansion in the horizontal and vertical directions, demonstrating a large negative Poisson's ratio value. In the close-up in Fig. 4d it is possible to note the scale interaction effect in the strongly deformed rank-two cuts intersecting the larger ones. The red circle in Fig. 4d highlights the large deformation of a rank-two cut, which constructively interacts with the deformations of the adjacent bigger cuts, enhancing the macroscopic auxetic effect.

**Experimental testing and validation**

Experiments were performed to validate the extreme auxetic behaviour of the described hierarchical plate. A thin 280.2 x 599.8 x 2 mm$^3$ PolyCarbonate (PC) specimen with the designed hierarchical distribution of traversing cuts was produced by a Roland EGX-600 professional engraving machine (accuracy 0.01 mm). The chosen geometry corresponds to a hierarchical ratio *s* = 1/4. The length of the rank-two cuts was reduced by 13% with respect to the geometry considered in Figs. 3 and 4, to avoid exceedingly thin connecting regions between cuts and high stress concentrations, and to make the experiments practically realizable. To mimic the real experiment, an additional numerical simulation of a two-dimensional model under plane stress conditions is performed in *Comsol Multiphysics* considering the same geometry of the specimen tested experimentally.

The specimen is mounted vertically in a loading frame (Messphysick MIDI 10) by uniformly clamping its upper and lower edges, and uniaxial tension is applied, as shown in Fig 5. During the test, the load is acquired with a DCSRC - 1 kN load cell and the displacement with a displacement transducer mounted in the loading frame. Small dots (0.5 mm in diameter) drawn at specific locations on the sample are used as tracking markers (also shown in Fig. 5c), and a 4K *Sony PXW-FS5 (3840 x 2160 pixel)* camera is used to track their displacements during loading (see movie in the Supplementary Material).

Results are shown in Fig. 5a: a vertical displacement is applied to the top of the specimen up to $u_{y,max}$ = 6 mm, and a perfectly linear elastic response is detected, leading to the estimation of an effective Young's modulus of 3.33 MPa. In the figure, the loading/unloading curve shows the perfectly linear elastic response of the specimen. To estimate the experimental average value of the Poisson's ratio, 8 markers are placed along the edges of a rectangular region in the central area of the hierarchical plate, as shown in Fig. 5c. The average horizontal and vertical displacements measured by using the 8 tracking points allow to estimate the Poisson's ratio that, after an initial variation due to specimen adjustment, tends to a value of $\bar{\nu} \cong -0.84$ (Fig. 5b). Such a value is in excellent agreement with the numerical value $\nu_{eff} = -0.86$, obtained with a finite numerical model analysed in *Comsol Multiphysics* and having the same geometry as the specimen tested in the lab.

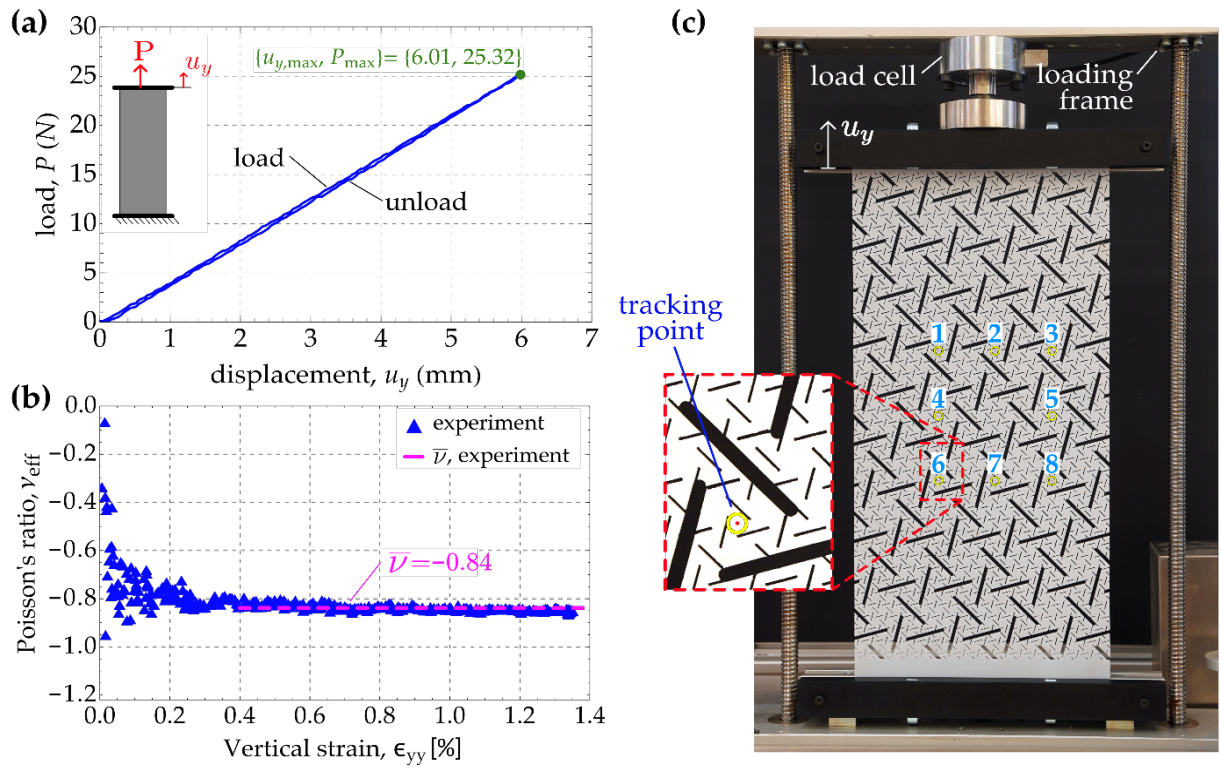

*Fig. 5: (a) Force vs displacement diagram (loading/unloading curve) recorded during the experimental test. (b) effective Poisson's ratio as a function of the vertical strain. The blue/triangle markers and the dashed/magenta line represent the experimental Poisson's ratio and its average value, respectively. (c) Experimental setup, with the specimen inside the loading frame. The tracking markers exploited to estimate the Poisson's ratio are highlighted and labelled from 1 to 8. An example of tracking marker is also highlighted in the inset.*

**Conclusions**

In conclusion, we have designed and tested a simple 2-D structure consisting of a hierarchical arrangement of rotated thin cuts, which displays extreme auxetic properties along each of its 6-fold symmetry directions, providing an isotropic architecture. The novelty of this structure is that it effectively expands uniformly in the plane when loaded uniaxially in any direction.

The design procedure starts from a previously considered single scale (non-hierarchical) design, adding smaller symmetrically rotated cuts, introducing an additional scale level into the structure. First, we show that this structure can be treated macroscopically as an isotropic homogeneous medium, with calculated effective quasi-static mechanical properties. Simulations show that these properties are susceptible to geometrical parameters, particularly the ratio between the lengths of the first- and second-rank unit cells. The designed structure enables a kirigami-like behaviour, and, exploiting the interaction between the scales, it is possible to achieve extreme isotropic auxetic properties. Numerical predictions are verified by experimental tests, with a large degree of confidence.

Future work will focus on a better understanding of the variation of effective mechanical properties of the proposed structures as a function of geometrical parameters, including the introduction of further hierarchical levels.

The designed auxetic structures can be useful for many applications in the biomedical field (e.g. for the design of novel types of stents), in the fabrication of impact protection devices and robust shock and sound absorbing materials, and in textile industry for the development of new fabrics.

**Acknowledgements**

MM, FB, ASG, DM and NMP are supported by the European Commission under the H2020 FET Open ("Boheme") grant No. 863179. MB acknowledges the financial support of Regione Autonoma della Sardegna, project ADVANCING.
**References**

[1] Evans K E (1991) Auxetic polymers: a new range of materials. *Endeavour* **15**(4) 170-174.

[2] Yeganeh-Haeri A, Weidner D J and Parise J B (1992) Elasticity of α-cristobalite: a silicon dioxide with a negative Poisson's ratio. *Science* **257** 650-652.

[3] Baughman R H, Shacklette J M, Zakhidev A A and Stafström S (1998) Negative Poisson's ratio as a common feature of cubic metals. *Nature* **392** 362-365.

[4] Grima J N, Jackson R, Alderson A and Evans K E (2000) Do zeolites have negative Poisson's ratios? *Adv. Mater.* **12** 1912-1918.

[5] Song F, Zhou J, Xu X, Xu Y and Bai Y (2008) Effect of a negative Poisson ratio in the tension of ceramics. *Phys. Rev. Lett.* **100** 245502.

[6] Lakes R S (1987) Foam structures with a negative Poisson's ratio. *Science* **235** 1038-1040.

[7] Lira C, Innocenti P and Scarpa F (2009) Transverse elastic shear of auxetic multi re-entrant honeycombs. *Compos. Struct.* **90**(3) 314-322.

[8] Li X, Wang Q, Yang Z and Lu Z (2019) Novel auxetic structures with enhanced mechanical properties. *Extreme Mech. Lett.* **27** 59-65.

[9] Peng X-L, Soyarslan C and Bargmann S (2020) Phase contrast mediated switch of

**Appendix A: Numerical verification of the isotropy of the hierarchical auxetic medium**

Hooke's law for a linear elastic isotropic medium under plane stress conditions can be expressed as follows:

$$\varepsilon_{xx} = \frac{\sigma_{xx} - \nu_{eff}\, \sigma_{yy}}{E_{eff}}, \qquad (A.1)$$

$$\varepsilon_{yy} = \frac{\sigma_{yy} - \nu_{eff}\, \sigma_{xx}}{E_{eff}}. \qquad (A.2)$$

Eqs. (A.1) and (A.2) can be solved in terms of the effective properties $E_{eff}$ and $\nu_{eff}$, yielding the expressions in Eqs. (2) and (3).

The isotropy of the solid phase and the six-fold symmetry of the geometry of the microstructure assure the isotropy of the effective constitutive properties. As an additional numerical check, we have performed three separate computations imposing three different macroscopic strain components, i.e. $\varepsilon_{xx}^{(1)} = 10^{-4}$, $\varepsilon_{yy}^{(2)} = 10^{-4}$ and $\varepsilon_{xy}^{(3)} = 10^{-4}$. The superscripts (1), (2) and (3) are introduced to distinguish the results of the three different computations.

We start from a generic orthotropic material under plane stress conditions, for which Hooke's law has the form

$$\varepsilon_{xx} = \frac{\sigma_{xx}}{E_{x,eff}} - \frac{\nu_{yx,eff}\, \sigma_{yy}}{E_{y,eff}}, \qquad (A.3)$$

$$\varepsilon_{yy} = \frac{\sigma_{yy}}{E_{y,eff}} - \frac{\nu_{xy,eff}\, \sigma_{xx}}{E_{x,eff}}, \qquad (A.4)$$

$$\varepsilon_{xy} = \frac{\sigma_{xy}}{2\, \mu_{xy,eff}}. \qquad (A.5)$$

From the first computation, where $\varepsilon_{xx}^{(1)} = 10^{-4}$ and $\varepsilon_{yy}^{(1)} = 0$, we obtain the stress components $\sigma_{xx}^{(1)}$ and $\sigma_{yy}^{(1)}$, which are inserted into Eqs. (A.3) and (A.4). Then, we repeat the same procedure for the second computation, where $\varepsilon_{xx}^{(2)} = 0$ and $\varepsilon_{yy}^{(2)} = 10^{-4}$. Successively, we solve the algebraic system of four equations, which yield

$$E_{x,eff} = \frac{\sigma_{xx}^{(1)} \sigma_{yy}^{(2)} - \sigma_{yy}^{(1)} \sigma_{xx}^{(2)}}{\varepsilon_{xx}^{(1)} \sigma_{yy}^{(2)}}, \quad (A.6)$$

$$E_{y,eff} = \frac{\sigma_{xx}^{(1)} \sigma_{yy}^{(2)} - \sigma_{yy}^{(1)} \sigma_{xx}^{(2)}}{\varepsilon_{yy}^{(2)} \sigma_{xx}^{(1)}}, \quad (A.7)$$

$$\nu_{xy,eff} = \frac{\varepsilon_{yy}^{(2)} \sigma_{yy}^{(1)}}{\varepsilon_{xx}^{(1)} \sigma_{yy}^{(2)}}, \quad (A.8)$$

$$\nu_{yx,eff} = \frac{\varepsilon_{xx}^{(1)} \sigma_{xx}^{(2)}}{\varepsilon_{yy}^{(2)} \sigma_{xx}^{(1)}}. \quad (A.9)$$

The results of the numerical simulations give $\sigma_{xx}^{(1)} = \sigma_{yy}^{(2)}$ and $\sigma_{yy}^{(1)} = \sigma_{xx}^{(2)}$ (see Tables A.1 and A.2), hence we retrieve $E_{x,eff} = E_{y,eff} = E_{eff}$ and $\nu_{xy,eff} = \nu_{yx,eff} = \nu_{eff}$, which can be computed using Eqs. (2) and (3), respectively. This restricts the orthotropic behaviour to a cubic one.

From the third simulation, where $\varepsilon_{xy}^{(3)} = 10^{-4}$, we determine $\sigma_{xy}^{(3)}$ and, using Eq. (A.5), we find $\mu_{xy,eff}$. Since $\mu_{xy,eff} = E_{eff}/\left(2(1 + \nu_{eff})\right)$, we verify that the medium is isotropic.

*Table A.1: Non-hierarchical geometry. Average values of stress and strain components and maximum value of von Mises stress obtained from three separate numerical computations.*

|      | Applied strains | | | Calculated stresses (Pa) | | | |
| --- | --- | --- | --- | --- | --- | --- | --- |
| Case | $\varepsilon_{xx}$ | $\varepsilon_{yy}$ | $\varepsilon_{xy}$ | $\sigma_{xx}$ | $\sigma_{yy}$ | $\sigma_{xy}$ | $s_{VM,max}$ |
| (1) | $10^{-4}$ | 0 | 0 | 28838 | -15894 | 0 | $0.723*10^6$ |
| (2) | 0 | $10^{-4}$ | 0 | -15894 | 28838 | 0 | $0.621*10^6$ |
| (3) | 0 | 0 | $10^{-4}$ | 0 | 0 | 44732 | $0.930*10^6$ |

In Tables A.1 and A.2 we report the average values of stresses and strains obtained from the numerical calculations for the non-hierarchical and the hierarchical geometries, respectively.

*Table A.2: Hierarchical geometry. Average values of stress and strain components and maximum value of von Mises stress from three separate numerical computations.*

|      | Applied strains | | | Calculated stresses (Pa) | | | |
|------|------|------|------|------|------|------|------|
| Case | $\varepsilon_{xx}$ | $\varepsilon_{yy}$ | $\varepsilon_{xy}$ | $\sigma_{xx}$ | $\sigma_{yy}$ | $\sigma_{xy}$ | $s_{VM,max}$ |
| (1) | $10^{-4}$ | 0 | 0 | 1860.0 | -1707.9 | 0 | $1.589*10^6$ |
| (2) | 0 | $10^{-4}$ | 0 | -1707.9 | 1860.0 | 0 | $1.403*10^6$ |
| (3) | 0 | 0 | $10^{-4}$ | 0 | 0 | 3567.9 | $2.236*10^6$ |

## Appendix B: Stress fields in the hierarchical and non-hierarchical geometry

In the following Figs. B.1 and B.2, we report the local stress fields components $s_{xx}$, $s_{yy}$ and $s_{xy}$ in the unit cell for the non-hierarchical and hierarchical geometries, respectively. Results correspond to applied macroscopic strains as in Fig. 2 and Tables A.1 and A.2.

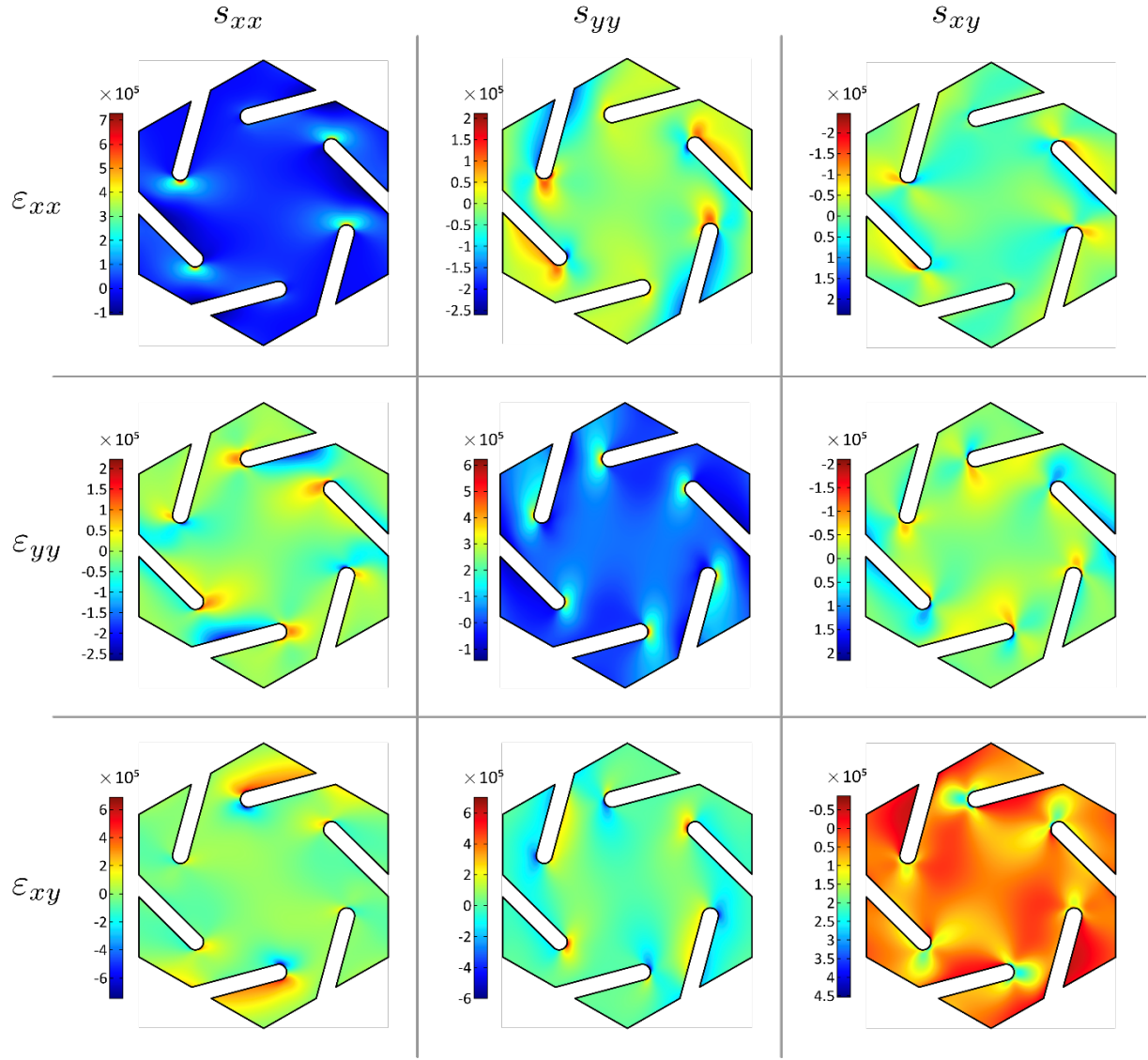

*Fig. B.1: Non-hierarchical geometry. Local stress fields in the unit cell for different applied average strains as in Fig. 2 and Tables A.1 and A.2.*

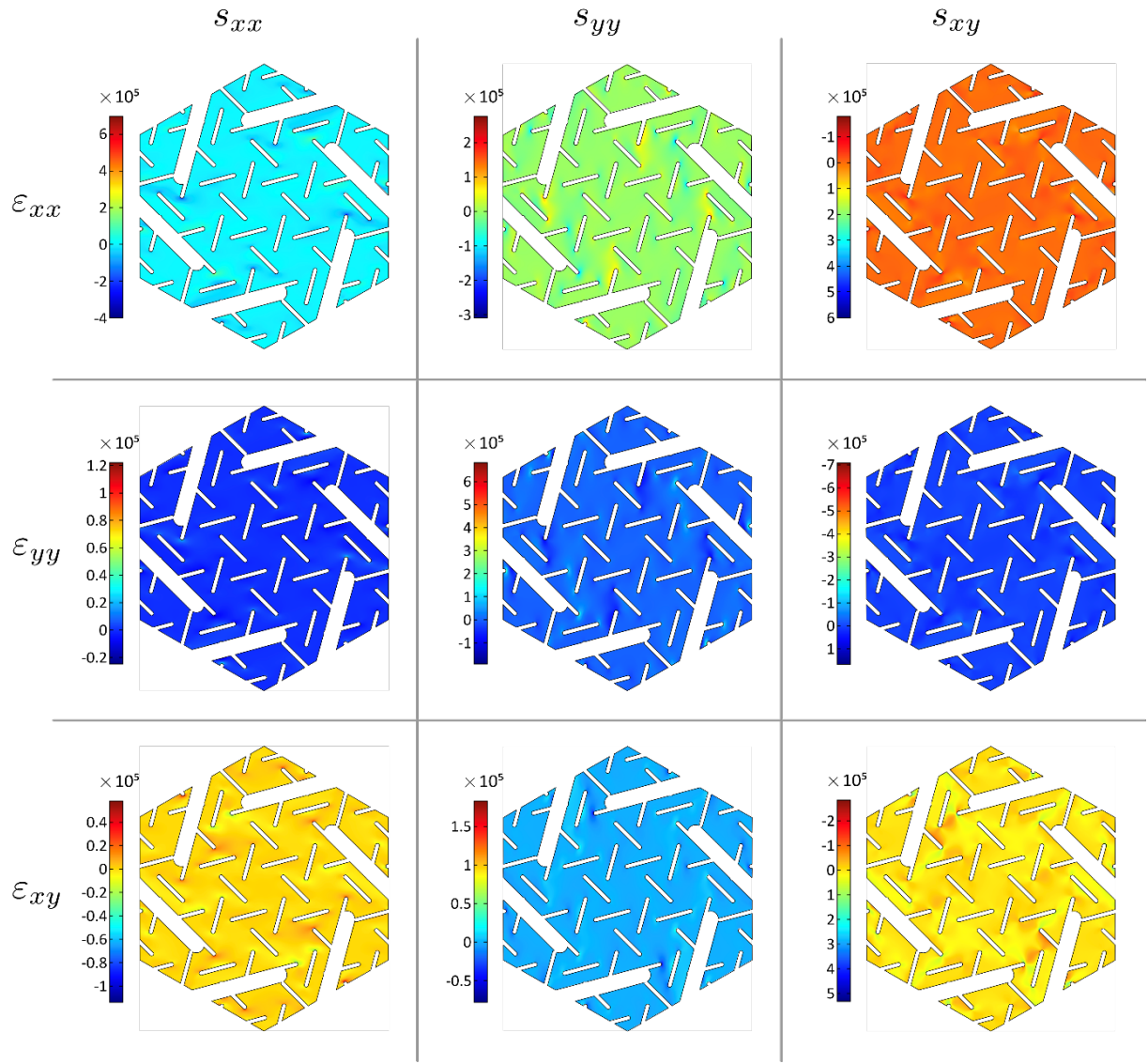

*Fig. B.2: Hierarchical geometry. Local stress fields in the unit cell for different applied average strains as in Fig. 2 and Tables A.1 and A.2.*